\documentclass[aps,twocolumn,floatfix, showkeys, superscriptaddress, nofootinbib]{revtex4}
\usepackage{amsmath}
\usepackage{graphicx,bm,hhline}
\usepackage{calrsfs}
\DeclareMathAlphabet{\pazocal}{OMS}{zplm}{m}{n}

\newcommand{\bcri}{{\bm R}_i}
\newcommand{\bk}{{\bm k}}
\newcommand{\bkp}{{\bm k}^\prime}

\newcommand{\br}{{\bm r}}

\newcommand{\nio}{n_I^0}

\newcommand{\wig}[1]{\mathrel{\hbox{\hbox to 0pt{\lower.6ex\hbox{$\sim$}\hss}\raise.4ex\hbox{$#1$}}}}

\newcommand{\nee}{N_e(\epsilon) }
\newcommand{\bp}{{\bm p}}
\newcommand{\bpp}{{\bp}^\prime}

\newcommand{\hatk}{\hat{\bk}}
\newcommand{\hatkp}{\hat{\bkp}}

\begin{document}
\title{Kubo-Greenwood approach to conductivity in dense plasmas with average atom models}
\author{C. E. Starrett}
\email{starrett@lanl.gov}
\affiliation{Los Alamos National Laboratory, P.O. Box 1663, Los Alamos, NM 87545, U.S.A.}

\date{\today}
\begin{abstract}
A new formulation of the Kubo-Greenwood conductivity for average atom models
is given.  The new formulation improves upon previous by explicitly including the
ionic-structure factor.  Calculations based on this new expression lead to much improved
agreement with {\it ab initio} results for DC conductivity of warm dense hydrogen and beryllium,
and for thermal conductivity of hydrogen.  We
also give and test a slightly modified Ziman-Evans formula for the resistivity that
includes a non-free electron density of states, thus removing an ambiguity
in the original Ziman-Evans formula.  Again results based on this expression are
in good agreement with {\it ab initio} simulations for warm dense beryllium and
hydrogen.  However, for both these expressions, calculations of the electrical
conductivity of warm dense aluminum lead to poor agreement at low temperatures compared
to {\it ab initio} simulations.
\end{abstract}
\maketitle

\section{Introduction}
An important aspect of modeling warm and hot dense matter is the calculation
of electron thermal and electrical conductivities.  The former
is of particular relevance in the field of inertial
confinement fusion \cite{lambert11, hu14} where it is the main
phenomena that determines the ablation of the cold deuterium/tritium fuel.  
Currently we have no reliable model that can predict accurate
thermal and electrical conductivities across
all temperature and density regimes of interest.  In particular,
as we move out of the degenerate electron regime the gold
standard method of Kohn-Sham density functional theory
molecular dynamics (KS-DFT-MD) coupled with the Kubo-Greenwood formalism
\cite{kubo57,greenwood58, hanson11, desjarlais02} quickly becomes computationally
prohibitive.  In the degenerate, or nearly degenerate regimes,
this method is thought to be accurate and agrees with 
experiments for materials under normal conditions \cite{sjostrom15}.

Average atom models provide an computationally efficient alternative 
at the cost of physical accuracy.  The central idea is that one tries
to calculate the properties of one atom in the plasma that is supposed
to represent the average of all atoms in the plasma.  Average atom models have been
used successfully for many years for equation of state calculations
\cite{feynman, liberman, piron3, wilson, scaalp, rozsnyai}.  They have also been used
for electrical conductivity calculations, primarily by coupling
to the Ziman-Evans (ZE) formula 
\cite{sterne07, perrot87, faussurier15, pain10, dharma06, rozsnyai08, burrill16}.
Recently, a systemic comparison of calculations of electrical conductivity 
using this method against Kubo-Greenwood KS-DFT-MD calculations \cite{burrill16} 
showed generally very good agreement between the methods provided that a judicious
choice was made when coupling the average atom model to the ZE 
formula.  However, the ZE formula, unlike the KG method, is
not easily generalized to thermal conductivity or optical conductivity.
The latter is useful as it can by used to calculate other optical
properties, including the opacity and reflectivity \cite{mazevet03}.

A formulation of the Kubo-Greenwood method for average atoms models has been 
developed by Johnson and co-workers \cite{johnson, johnson09, johnson2}.  However, a subsequent systematic
analysis of the method compared to KS-DFT-MD showed
some serious inaccuracies \cite{starrett12a}.  Unlike the ZE
formulation, Johnson's KG formulation does make not explicit account
of the ion-ion structure factor $S(k)$.  In this work, we give an
alternative derivation of the KG formulation for average atom models
that explicitly accounts for $S(k)$.  The new formulation recovers
Johnson's result when $S(k)=1$.  We also give the equations for
thermal and optical conductivity.

To evaluate this new formulation we make comparisons to KS-DFT-MD
calculations for hydrogen \cite{lambert11} and beryllium \cite{hanson11}.
We also compare to other models \cite{faussurier15, sjostrom15} and
experiments for aluminum \cite{milchberg88, sperling15}.  We use
the recently developed pseudo-atom molecular dynamics (PAMD)
\cite{starrett15, starrett13}
to generate the necessary inputs for the KG equation.

In addition to this, we present a slightly modified Ziman-Evans
formula that takes into account a non-free electron
density of states (DOS).  The original ZE formula assumes
a free electron DOS and this leads to an ambiguity
in the choice of chemical potential and density of
scattering electrons.  This point was discussed in
detail in \cite{burrill16}.  The present reformulation
recovers the original form of the ZE equation when
the DOS goes to the free electron form and removes
the ambiguity when the DOS is not free electron like.
We compare calculations based on this new ZE formulation
to the new KG formulation and to the KS-DFT-MD results.

The structure of this paper is as follows.  In section \ref{sec_kg}
we derive the Kubo-Greenwood expression for average atom 
models with explicit account of the ion-ion structure
factor.  We also give the expression for the thermal
conductivity.  In section \ref{sec_ze} we show how the
Ziman-Evans formula for the inverse resistivity is modified
to account for a non-free electron density of states.
In section \ref{sec_pamd} we discuss the connection of these
formulas to the the Pseudo-Atom Molecular Dynamics (PAMD)
average atom model. In section \ref{sec_res} we use the PAMD model
with the new KG and ZE expressions to calculate the DC
electrical conductivity of warm dense hydrogen, beryllium
and aluminum, and compare to available simulations, models
and experiments.  For hydrogen we also compare thermal
conductivity calculations to KS-DFT-MD simulation results.
Lastly, in section \ref{sec_con} we draw our conclusions.
Throughout we use Hartree atomic units 
in which $\hbar = m_e = e = 1$.

\section{Kubo-Greenwood approximation\label{sec_kg}}
The Kubo-Greenwood expression for the conductivity is \cite{greenwood58}
\begin{eqnarray}
\sigma(\omega) & = &\frac{-2\pi}{V} \int\,d\epsilon\, 
\frac{f(\epsilon_m) - f(\epsilon_n)}{\omega}
\int\,d^3k_m\,
\int\,d^3k_n\,\nonumber\\
&& \times
 \left < \left| J_{mn} \right|^2 \right> 
 \delta(\epsilon_m - \epsilon - \omega)
\delta(\epsilon_n - \epsilon)
\end{eqnarray}
with
\begin{equation}
 J_{mn} \equiv \int_V\,d^3r\,
\psi_{{\bm k}_m}^*(\bm r) \hat{\bm v}_z
\psi_{{\bm k}_n}(\bm r) 
\end{equation}
where $\epsilon_{n(m)} = k_{n(m)}^2 / 2$ is the energy of the initial (final) electron state and $\psi_{{\bm k}_{n(m)}}(\br)$
is the corresponding wave function, $f(\epsilon)$ is the Fermi-Dirac occupation factor and $\hat{\bm v}_z$ is
the velocity operator in the $\hat{\bm z}$ direction.
Following Evans \cite{evans73}, we now assume that the potential felt by a electron is of muffin-tin form.  In this widely
used approximation the total scattering potential is the sum of non-overlapping potentials,
centered on each nuclear site.  Each muffin-tin potential is contained in a sphere of volume $V_{MT}$. 
Again following Evans \cite{evans73} we further assume that the wave function inside each sphere
the wave function is given by
\begin{equation}
\psi_{{\bm k}_n}(\bm r) = 
\bar{\psi}_{{\bm k}_n}(\bm r) e^{\imath {\bm k}_n \cdot {\bm R}_\alpha }
\end{equation}
where
\begin{equation}
\bar{\psi}_{{\bm k}_n}(\bm r) =
\sum_l\sum_m \imath^l e^{\imath \delta_l(k_n) }
Y_{lm}(\hat{{\bm k}_n}) 
Y_{lm}^*(\hat{{\bm r}_n}) 
\frac{y_{l}(r,k_n)}{r\sqrt{k}} 
\label{psibar}
\end{equation}
Here ${\bm R}_\alpha$ is the position vector of nucleus $\alpha$.  Further assuming that 
each muffin tin potential is identical and using the definition of the
ion-ion structure factor:
\begin{equation}
S(k) = \frac{1}{N} \left< \rho_{\bk} \rho_{-{\bk}} \right>
\end{equation}
where 
\begin{equation}
\rho_{\bm k} = \sum_{\alpha=1}^{N}
e^{\imath {\bm k} \cdot {\bm R}_\alpha }
\end{equation}
the Kubo-Greenwood conductivity expression is reduced to 
\begin{eqnarray}
\sigma(\omega) & = & -2\pi n_I^0 \int\,d\epsilon\, 
\frac{f(\epsilon_m) - f(\epsilon_m)}{\omega}
\int\,d\hat{\bk}_m\,
\int\,d\hat{\bk}_n\, \nonumber\\ 
&& \times 
|\sqrt{k_n\,k_m}\bar{J}_{mn} |^2
S(|{\bk_n}-{\bk_m}|)
\label{sigjbar}
\end{eqnarray}
where $\epsilon_m = \epsilon + \omega$, $\epsilon_n = \epsilon$, $n_I^0 = N/V$
\begin{equation}
\bar{J}_{mn} \equiv \int_{V_{MT}}\,d^3r\,
\bar{\psi}_{{\bm k}_m}^*(\bm r) \hat{\bm v}_z
\bar{\psi}_{{\bm k}_n}(\bm r) 
\end{equation}
Using equation (\ref{psibar}) in (\ref{sigjbar}) and after some lengthy
algebra (see appendix) we arrive at the result 
\begin{equation}
\sigma(\omega) = 
\sigma^{(1)}(\omega) +
\sigma^{(2)}(\omega) +
\sigma^{(3)}(\omega)
\label{sigkg}
\end{equation}
with
\begin{eqnarray}
\sigma^{(1)}(\omega) & = & -2\pi n_I^0 \int\,d\epsilon\, 
\frac{f(\epsilon_m) - f(\epsilon_m)}{\omega}
\nonumber\\ 
&& 
\!  \!  \!  \!  \!  \!  \!  \!  \!  \!  \!  \!
\!  \!  \!  \!  \!  \!  \!  \!  \!  \!  \!  \!
\times 
\sum_{l=0}^\infty |I_A(l)|^2
\frac{(l+1)}{2(2l+3)}
\left[
(1+\frac{l}{2})I_s^{(1)} + \frac{l}{2}I_s^{(2)}
\right]
\end{eqnarray}
\begin{eqnarray}
\sigma^{(2)}(\omega) & = & 2\pi n_I^0 \int\,d\epsilon\, 
\frac{f(\epsilon_m) - f(\epsilon_m)}{\omega}
\nonumber\\ 
&& 
\!  \!  \!  \!  \!  \!  \!  \!  \!  \!  \!  \!
\!  \!  \!  \!  \!  \!  \!  \!  \!  \!  \!  \!
\times 
\sum_{l=0}^\infty I_A(l) I_B(l+2)
\cos(\delta_l(k_n) - \delta_{l+2}(k_n)) \nonumber \\
&&
\!  \!  \!  \!  \!  \!  \!  \!  \!  \!  \!  \!
\!  \!  \!  \!  \!  \!  \!  \!  \!  \!  \!  \!
\times 
\frac{(l+1)(l+2)}{2(2l+3)}
\left[
3\, I_s^{(2)} - I_s^{(1)}
\right]
\end{eqnarray}
\begin{eqnarray}
\sigma^{(3)}(\omega) & = & -2\pi n_I^0 \int\,d\epsilon\, 
\frac{f(\epsilon_m) - f(\epsilon_m)}{\omega}
\nonumber\\ 
&& 
\!  \!  \!  \!  \!  \!  \!  \!  \!  \!  \!  \!
\!  \!  \!  \!  \!  \!  \!  \!  \!  \!  \!  \!
\!  \!  \!  \!  \!  \!  \!  \!  \!  \!  \!  \!
\times 
\sum_{l=0}^\infty |I_B(l)|^2
\frac{l}{4(2l-1)}
\left[
(l-1) I_s^{(1)} + (l+1) I_s^{(2)}
\right]
\end{eqnarray}
where
\begin{equation}
I_s^{(1)} \equiv \int_{-1}^{1} dx\, S(\sqrt{(k_n^2 + k_m^2 -2 k_n k_m\, x)})
\end{equation}
\begin{equation}
I_s^{(2)} \equiv \int_{-1}^{1} dx\,x^2\, S(\sqrt{(k_n^2 + k_m^2 -2 k_n k_m\, x)})
\end{equation}
and $I_A$ and $I_B$ are the same as in reference \cite{starrett12a},
\begin{eqnarray}
I_A(l, k_n, k_m) & \equiv & \int\limits_0^{R_{MT}} dr\, y_{l+1}(r,k_m)\nonumber\\
&& \!\!\!\!\!\!\!\!\!\!\!\times 
\left(
\frac{\partial y_{l_i}(r,k_n)}{\partial r} - (l + 1) \frac{y_{l}(r,k_n)}{r}
\right)
\label{eqn_ia}
\end{eqnarray}
\begin{eqnarray}
I_B(l, k_n, k_m) & \equiv & \int\limits_0^{R_{MT}} dr\, y_{l-1}(r,k_m)\nonumber\\
&& \!\!\!\!\!\!\!\!\!\!\!\times 
\left(
\frac{\partial y_{l_i}(r,k_n)}{\partial r} + l \frac{y_{l}(r,k_n)}{r}
\right)
\label{eqn_ib}
\end{eqnarray}

In the limit $S(k) \to 1$  $\forall k$, $\sigma^{(2)}(\omega) = 0 $ and
the expression for the conductivity is reduced to that of Johnson's result
\cite{johnson, starrett12a} provided that the integral over the muffin tin volume $V^{MT} = 4/3\pi R_{MT}^3$
is instead taken over all space.  We return to this point in section \ref{sec_pamd}

As shown in reference \cite{starrett12a} the thermal conductivity $\kappa$ can be calculated in a straight forward extension.
For a plasma of temperature $T$
\begin{equation}
\kappa = \frac{1}{T} \left( 
{\cal L}_{22} - \frac{{\cal L}_{12}^2}{{\cal L}_{11}}
\right)
\end{equation}
where
\begin{eqnarray}
{\cal L}_{nm} = 
& = & 
(-1)^{n+m} 2\pi n_I^0 
\int\,d\epsilon\, (\epsilon -\mu_e)^{n+m-2}\nonumber\\
&& \times
\frac{\partial f(\epsilon)}{ \partial \epsilon}
\int\,d\hat{\bk}_m\,
\int\,d\hat{\bk}_n\, \nonumber\\ 
&& \times 
|\sqrt{k_n\,k_m}\bar{J}_{mn} |^2
S(|{\bk_n}-{\bk_m}|)
\end{eqnarray}
Clearly, $\sigma_{DC} = \sigma(0) = {\cal L}_{11}$.  The Lorenz number is defined as
\begin{equation}
L = \frac{\kappa}{T \sigma_{DC}}
\end{equation}
For fully degenerate electrons this takes the value $L=\pi^2/3$, while
for fully non-degenerate electrons this takes the value 1.597.
Lastly we note that the extension of this formulation to mixtures
is straightforward, but is not explored here.

\section{Ziman-Evans expression with an explicit density of states\label{sec_ze}}
The Ziman-Evans expression for the inverse conductivity $R$ is \cite{ziman61, evans73}
\begin{equation}
R  = \frac{1}{\sigma_{DC}} = -\frac{1}{3 \pi^2 (n_e^*)^2} \int_0^\infty d\epsilon \frac{df}{d\epsilon} v^3 \frac{1}{\tau_p}
\label{inverse}
\end{equation}
where $n_e^*$ is the density of scattering electrons and $f(\epsilon,\mu_e^*)$ is the Fermi-Dirac occupation factor
and we have expanded the notation to indicate that it depends on the chemical potential $\mu_e^*$.
The relaxation time $\tau_p$ is defined in terms of the generalized momentum transport cross section
\begin{equation}
\sigma_{TR}(p) = 2\pi \int_0^\pi d\theta |T_{\bp \bpp}|^2 (1-\cos\theta)\sin\theta
\end{equation}
and the transition matrix element $T_{\bp \bpp}$ is given by \cite{evans73}
\begin{equation}
|T_{\bp \bpp}|^2 = \frac{d\sigma}{d\theta}(p,\theta) S(q)
\end{equation}
where $q^2 = 2p^2(1-\cos\theta)$, and $S(q)$ is the ion-ion structure factor and $d\sigma/d\theta$ is the differential scattering
cross section \cite{burrill16}.  $\bp$ ($\bpp$) is the initial (final) momentum of the electron before (after)
the collision with the atom.  Only elastic collisions are included, in which $|\bp| = |\bpp|$.  
The final result for the relaxation time is
\begin{equation}
\frac{1}{\tau_p} = \pi n_I^0 \frac{v}{p^4}\, 
\int_0^{2p} dq\, q^3 \frac{d\sigma}{d\theta}(p,\theta) S(q)
\end{equation}
where $p=m_e v = \sqrt{2 \epsilon}$.

As explained in reference \cite{burrill16} a challenge when using equation (\ref{inverse}) is
the ambiguity in what one should choose for $n_e^*$ and $\mu_e^*$.  This is due to the implicit 
free-electron density of states.  
Equation (\ref{inverse}) is modified to include a non-free electron density of states by introducing a factor
$\nee$ into the integrand \cite{potekhin96}:
\begin{equation}
R     =   \frac{1}{(n_e^*)^2}  \int_0^\infty d\epsilon \left(-\frac{df}{d\epsilon}\right) \nee \frac{1}{\tau_p}
\label{r_dos}
\end{equation}
where 
\begin{equation}
\nee = n_I^0 \int_0^\epsilon d\epsilon^\prime \chi(\epsilon^\prime) 
\label{nee}
\end{equation}
$\chi(\epsilon)$ is the density of states such that the number of valence electrons per atom
$\bar{Z}$ is
\begin{equation}
\bar{Z} = \frac{\bar{n}_e^0}{n_I^0} = \int_0^\infty d\epsilon \chi(\epsilon)  f(\epsilon, \mu_e)
\label{zbar}
\end{equation}
Here $\mu_e$ is the calculated, ``physical'' chemical potential and $\bar{n}_e^0$ is the density of
valence electrons.  Thus we can identify $n_e^* = \bar{n}_e^0$ and $\mu_e^* = \mu_e$.  Such a choice
cannot be used with equation (\ref{inverse}) without introducing an inconsistency, and one is
forced to compromise (see \cite{burrill16}).
In the limit as $T \to 0$ in equation (\ref{r_dos}) we recover the expected Drude form
\begin{equation}
\frac{1}{R}  = \bar{n}_e^0\, \tau_{v_F} 
\label{drude}
\end{equation}
The usual expression for the resistivity (\ref{inverse}) is recovered from equation (\ref{r_dos})
if we take the density of states to be its free electron form
\begin{equation}
\chi^{free}(\epsilon) = \frac{\sqrt{2 \epsilon}}{n_I^0 \pi^2 }
\end{equation}
then
\begin{equation}
N_e^{free}(\epsilon) = \frac{ v^3 }{3 \pi^2 }
\end{equation}

\section{Connection to the PAMD model\label{sec_pamd}}
The above formulations can be used with any model that give access to the electron
scattering potential $V^{scatt}$ and the ion-ion structure factor $S(k)$.  From
the scattering potential one can determine the chemical potential $\mu_e$,
the wavefunctions $y_l(r,k)$, and the density of states $\chi(\epsilon)$.
As in reference \cite{burrill16} we shall use the pseudo-atom molecular
dynamics (PAMD) \cite{starrett15} model to generate these inputs. 
This model had been described in detail elsewhere \cite{starrett15, starrett13, starrett14}
and has successfully been used to calculate equation of state, ionic structure
and ionic transport \cite{daligault16} in the warm and hot dense matter
regime.  Its connection with the ZE equations was explored in detail in
\cite{burrill16}.  In short, in PAMD the electronic structure of one
pseudo-atom in a plasma is calculated using density functional theory (here
we restrict ourselves to Kohn-Sham DFT\footnote{For the results in section \ref{sec_res} we
have used the zero temperature Local Density Approximation \cite{dirac} for aluminum and the
finite temperature LDA for beryllium and hydrogen \cite{ksdt}.}).  By coupling this electronic structure
to the integral equations of fluid theory (the quantum Ornstein-Zernike (QOZ) equations)
one calculates a parameter-free ion-ion pair interaction potential.  This can be used
in molecular dynamics simulations or, as we have here and in reference \cite{burrill16},
directly in the QOZ equations to generate $S(k)$.

As discussed in reference \cite{burrill16} there are at least two reasonable choices 
for the scattering potential: the pseudo-atom potential $V^{PA}(r)$ and the average
atom potential $V^{AA}(r)$.  The former can be used to construct the total
scattering potential of the plasma when combined with a set on nuclear position
vectors $\{{\bm R}_i\}$
\begin{equation}
V^{tot}(\br) = \sum_i V^{PA}(|\br - \bcri|)
\label{vsuper}
\end{equation}
while the latter $V^{AA}$ is a muffin-tin like potential that extends to the ion-sphere
radius.  Given the above derivation of the KG formula one would expect
$V^{AA}$ to be the most reasonable choice.  However, it was found in \cite{burrill16}
that when using the (similarly derived) ZE formulation $V^{PA}$ gives generally much better agreement with KS-DFT-MD results that
are thought to be accurate.  This surprising result was explained by realizing that
in the Born limit of scattering (i.e. where the scattered state is a plane wave)
the total scattering cross section separates into a sum over cross sections
for each atom.  
In that limit the total scattering cross section is given by the Fourier transform
of $V^{tot}$.  Thus to recover the Born limit it is necessary to use the $V^{PA}$
potential.  Though this choice violates the condition of non-overlapping
muffin-tins, it compensates for the neglect of multiple scattering effects in our
wavefunctions by essentially treating them in the Born approximation, while
allowing strong single site scattering though the use of the t-matrix approximation.

In the KG formulation we similarly neglect multiple scattering contributions and we
might expect the same potential ($V^{PA}$) to also lead to improved results, again
letting the muffin tin volume being replaced by a integral over all space.  We note
that in Johnson's formulation \cite{johnson}, where $S(k)= 1$, the $V^{AA}$ potential is used and
the integral is taken over all space.  There, the physical model is an average atom model
in jellium.  One could restrict the integrals to be inside the muffin-tin (ion-sphere)
volume only, but this results in conductivities that are orders of magnitude
incorrect!  The reason for this gross error was first suggested in reference \cite{dulca00}.
In that work the isolated cluster method (ICM) was use to calculate conductivity of
clusters of atoms with up to 201 atoms using the real-space Green's function method.
The average atom model is essentially an ICM with one atom \cite{starrett15a}.  
In reference \cite{dulca00} they found
that the conductivity was dependent on the size of the cluster and that convergence was not achieved
with up to 201 atoms.  The explanation given was that the cluster approach can only yield correct
results if the mean free path of the electron is approximately equal to, or smaller than,
the size of the cluster -- if the mean free path of the electron is larger than the the cluster,
the scattering processes that lead to a finite conductivity cannot be expected to be accurately
modeled in the cluster representation \cite{dulca00}.  Clearly, here with our neglect of multiple
scattering effects, our cluster size is effectively the volume of the ion-sphere, hence unless the
resistivity is very large (corresponding to very small mean free paths) the results will
be substantially incorrect.  By letting the integration volume go to infinity this obviously
corrects for this defect.  However the price is that, since no other scattering events 
are included in the electronic wavefunction, the calculated conductivity diverges 
as $\omega^{-2}$ \cite{johnson}!
Johnson introduced a practical method to correct for this divergence: the optical
conductivity is required to have a Drude like behavior by multiplying by
$\omega^2/(\gamma^2 + \omega^2)$.  $\gamma$ is found by enforcing the frequency
sum rule
\begin{equation}
\int\limits_0^\infty d\omega \sigma(\omega) = \frac{\nio \pi \bar{Z}}{2} \label{gam_cond}
\end{equation}
where $\bar{Z}$ is the number of valence electrons per ion.  We note that we are only 
considering the free-free contribution to the conductivity here.  
This combination of letting the integration volume go to infinity and enforcing
the Drude form represents a reasonable way of capturing the main effects
important to the calculation of the conductivity, but clearly introduces
a significant source of uncertainty into the quality of the calculation.

In the next section we present calculations using the KG equation (\ref{sigkg}) using
this infinite volume correction.  We show results using the two potentials 
$V^{PA}$ and $V^{AA}$.  For $V^{AA}$ we also show the result with $S(k)=1$,
which is the same as Johnson's formulation \cite{johnson}, these
can be compared to the nearly identical calculations presented in
reference \cite{starrett12a}.  We show results from the Ziman-Evans
equation with the non-free density of states included, equation (\ref{r_dos}),
that we label ZE-DOS.  Our notation is summarized in table \ref{tab_note}.

\begin{table*}
\begin{center}
\begin{tabular}{ l l }
\hline
\hline
 Notation & Description \\
\hline
 KG: $V^{PA}$ &  Equation (\ref{sigkg}) with wave functions calculated in the $V^{PA}(r)$ potential.  \\
 KG: $V^{AA}$ &  Equation (\ref{sigkg}) with wave functions calculated in the $V^{AA}(r)$ potential.  \\
 KG: $V^{AA}$, $S(k)=1$ &  Same as KG: $V^{AA}$, but with $S(k)=1$.  This is equivalent to Johnson's formulation \cite{johnson}.  \\
\hline
 ZE-DOS: $V^{AA}$ or $V^{PA}$  & Equation (\ref{r_dos}) with either potential.  \\
 ZE-$\mu_e$   & Equation (\ref{inverse}) with the ``physical" chemical potential (see \cite{burrill16}).  \\
 ZE-$\bar{\mu}_e$   & Equation (\ref{inverse}) with the free electron chemical potential (see \cite{burrill16}).  \\
\hline
\hline
\end{tabular}
\caption{\label{tab_note} Descriptions of notation used in the figures.
}
\end{center}
\end{table*}

\begin{figure}
\begin{center}
\includegraphics[scale=0.4]{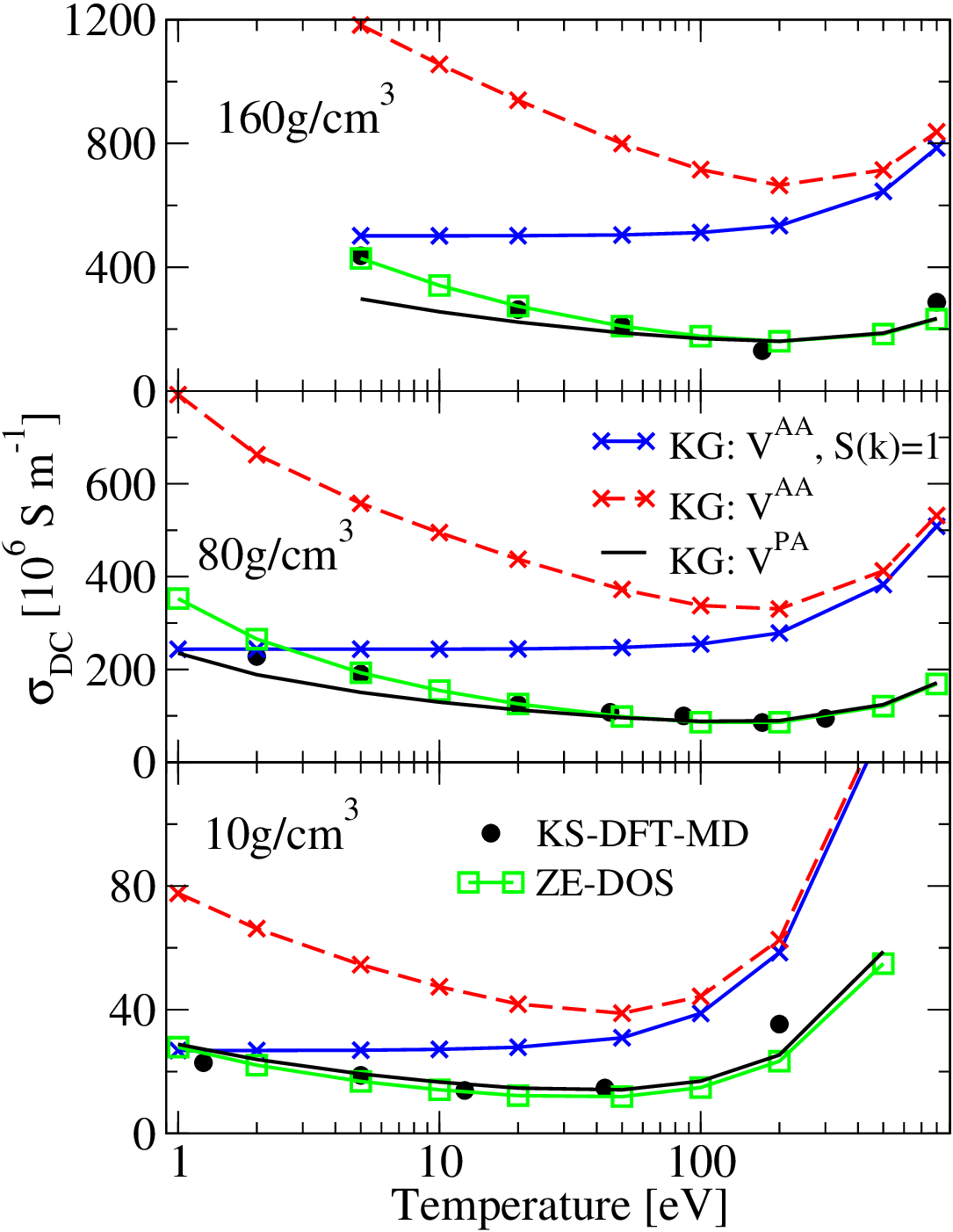}
\end{center}
\caption{(Color online) Electrical conductivity of dense hydrogen. 
The KS-DFT-MD results are from reference \cite{lambert11}. 
}
\label{fig_h1}
\end{figure}

\begin{figure}
\begin{center}
\includegraphics[scale=0.4]{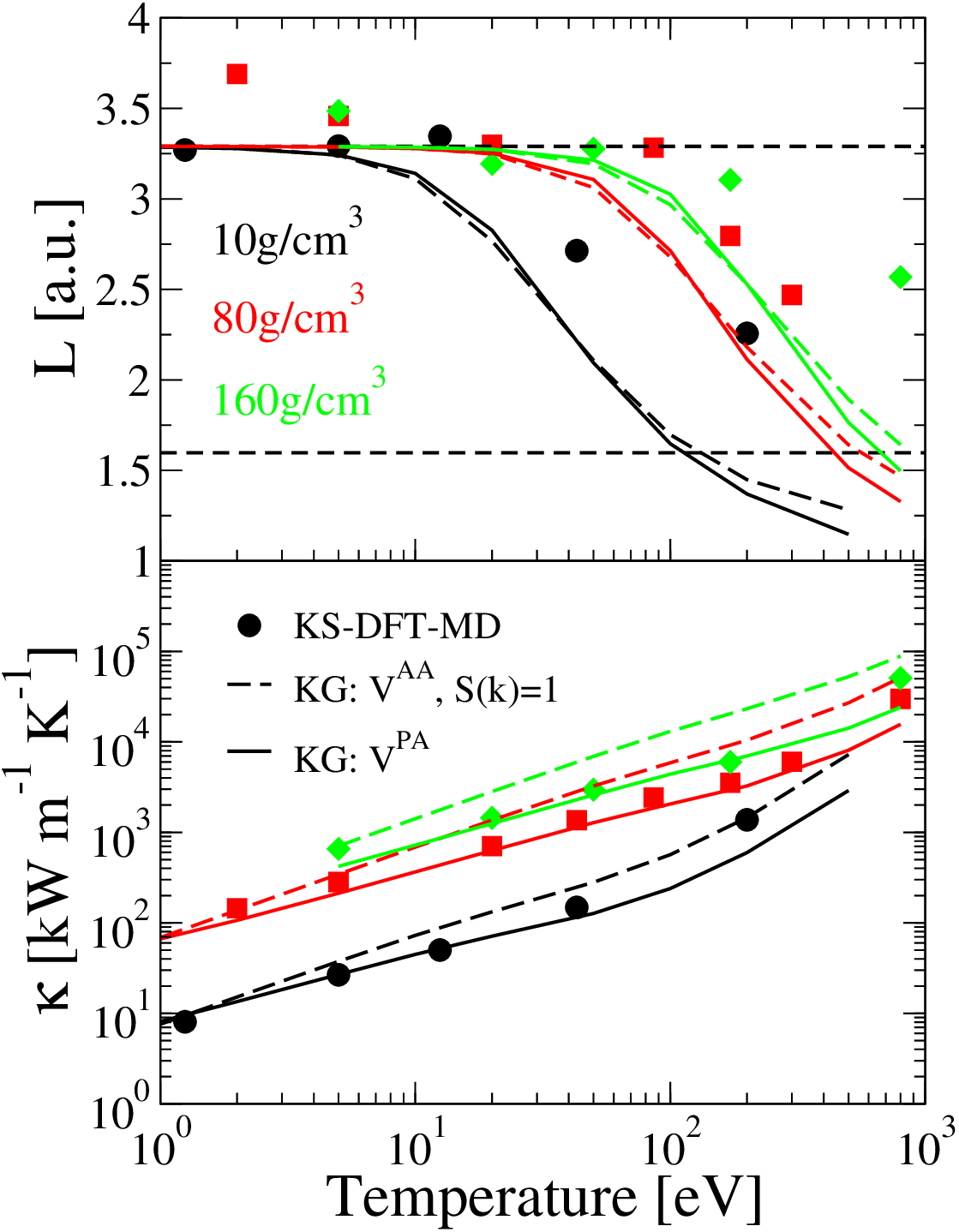}
\end{center}
\caption{(Color online) Lorenz number and thermal conductivity of dense hydrogen.  
The KS-DFT-MD results are from reference \cite{lambert11}.  The horizontal dashed lines
for the Lorenz number indicate the degenerate and non-degenerate limits.
}
\label{fig_hthermal}
\end{figure}

\begin{figure}
\begin{center}
\includegraphics[scale=0.35]{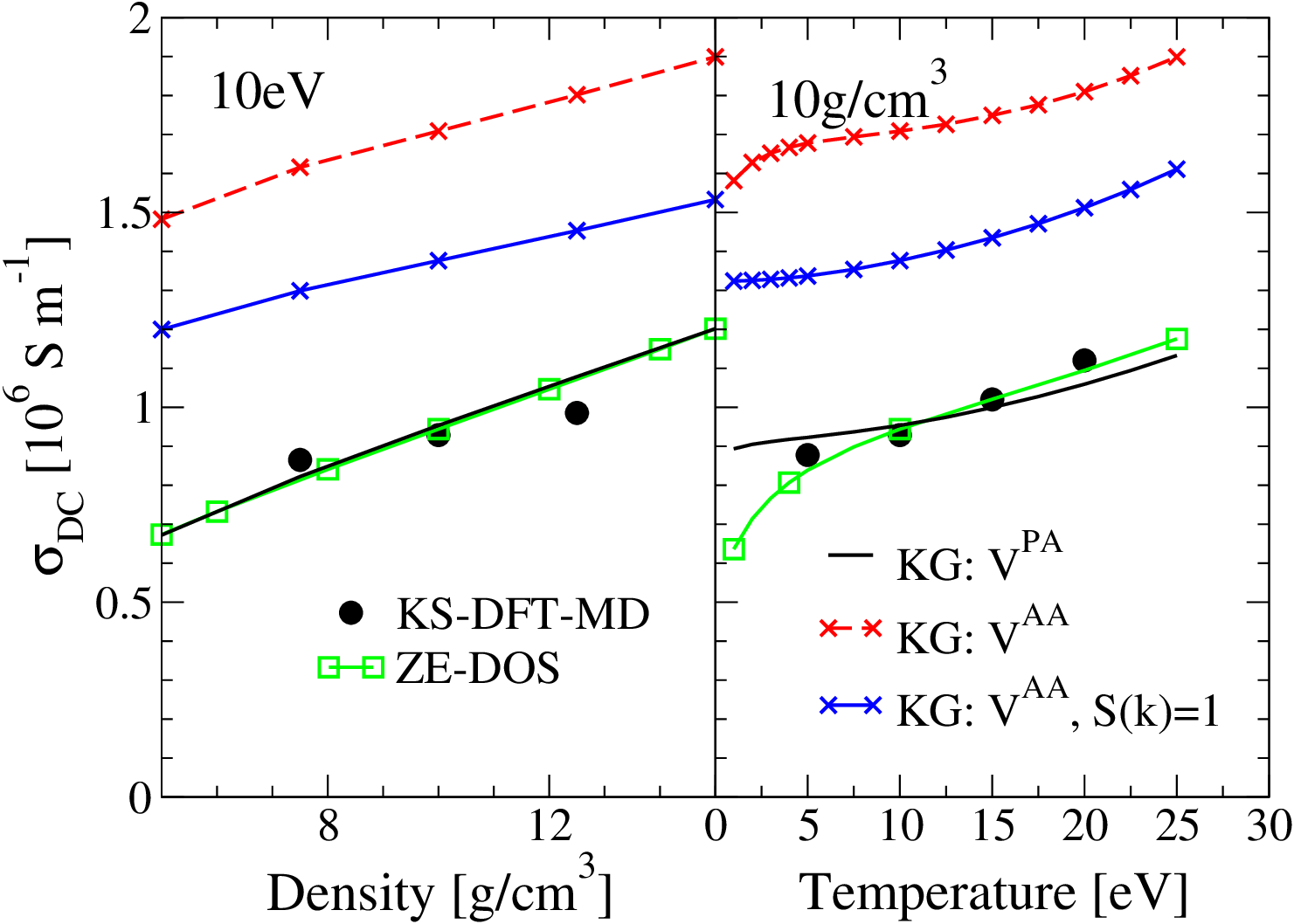}
\end{center}
\caption{(Color online) Electrical conductivity of dense beryllium.  
The KS-DFT-MD results are from references \cite{starrett12a, hanson11}. 
}
\label{fig_be1}
\end{figure}

\begin{figure}
\begin{center}
\includegraphics[scale=0.4]{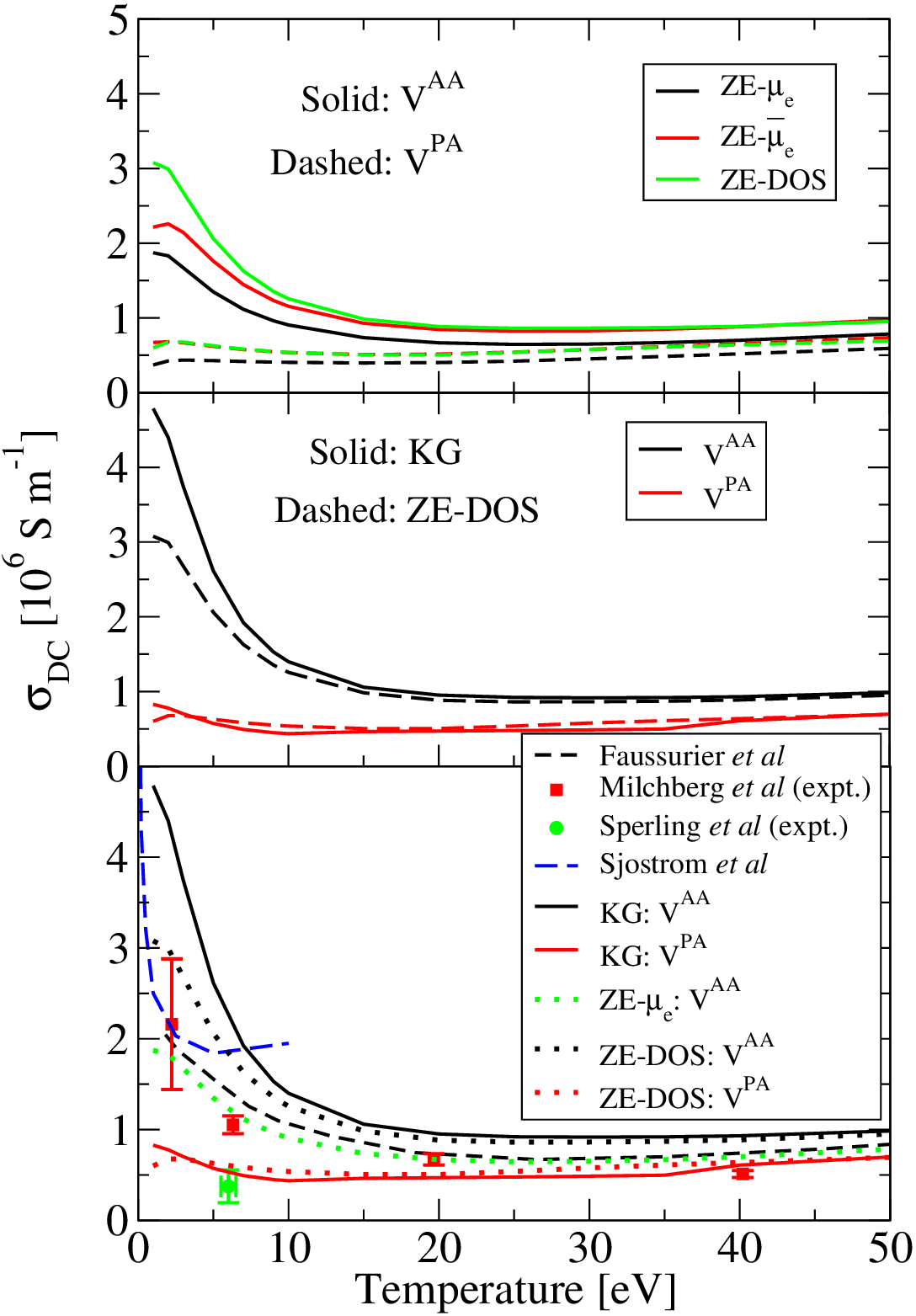}
\end{center}
\caption{(Color online) Electrical conductivity of solid density (2.7 g/cm$^3$) aluminum.  
}
\label{fig_alsigff}
\end{figure}

\section{Results\label{sec_res}}
In figure \ref{fig_h1} we show results for the DC conductivity $\sigma_{DC}$ for dense hydrogen.  We compare to
KS-DFT-MD results from \cite{lambert11}.  Clearly both the Ziman-Evans (ZE-DOS) calculations based on equation
(\ref{r_dos}) and the Kubo-Greenwood (KG) calculations based on equation (\ref{sigkg}) agree well 
with the KS-DFT-MD results when the pseudoatom potential ($V^{PA}$) is used.  
A comparison of the ZE-DOS results to those presented in reference \cite{burrill16} with the two choices of 
chemical potential (not shown, but compare to results in reference \cite{burrill16}) reveals little effect 
from the new formulation, as we would expect, since the two results in \cite{burrill16} are similar.
Where a difference is seen the ZE-DOS calculations most closely follows the calculations based on
the choice $\bar{\mu}_e$.  KG calculations
using $V^{AA}$ are also shown in figure \ref{fig_h1}.  When the structure factor is included the qualitative agreement is
reasonable but the quantitative agreement is poor.  Without the structure factor, even the
qualitative agreement is poor and the results agree with those presented in \cite{starrett12a}.
Therefore the combination of the use of $V^{PA}$ with the inclusion of the ionic
structure factor are both vital to find agreement with the KS-DFT-MD results in this case.

In figure \ref{fig_hthermal} we show calculations of the thermal conductivity $\kappa$ and
Lorenz number $L$ for the same dense hydrogen conditions as in figure \ref{fig_h1}.
We show calculations using the average atom potential $V^{AA}$ with $S(k)=1$, 
and using $V^{PA}$.  
The results using $V^{AA}$ amount to essentially the same calculations as 
in \cite{starrett12a}, and the results are very similar, i.e. generally the thermal
conductivity overestimates the KS-DFT-MD and the trends are only broadly in agreement.
In contrast, the results using $V^{PA}$ agree well with KS-DFT-MD, though some differences
at high temperature are found.  For the Lorenz number using either potential leads to similar
results, and both agree with the degenerate limit at low temperature, and 
tend to underestimate KS-DFT-MD at high temperature.

In figure \ref{fig_be1} we show results for warm dense beryllium.  As for hydrogen, using
the $V^{AA}$ leads to an overestimation of the KS-DFT-MD results, and using $V^{PA}$ with
$S(k)$ from PAMD leads to very good agreement, using either the ZE-DOS or the KG methods.
These results, figures \ref{fig_h1} to \ref{fig_be1}, demonstrate that both models introduced
in this work can successfully predict conductivities in warm dense matter.   For clarity of the figures,
we have not shown an explicit comparison for ZE-DOS to the results presented in \cite{burrill16} which used
the original ZE equation (\ref{inverse}) with the two choices of chemical potential (ZE-$\mu_e$ and 
ZE-$\bar{\mu}_e$). However, a comparison of the results reveals that ZE-DOS generally lies between
the two and closer to ZE-$\bar{\mu}_e$ than ZE-$\mu_e$.

In figure \ref{fig_alsigff} we show results for warm dense aluminum.  Here the situation
is much more complicated.  In the top panel we show calculations based on the
Ziman-Evans formula only.  We compare both potentials ($V^{AA}$ and $V^{PA}$) as well
as the new ZE-DOS formulation to original ZE formulation with the two choices of chemical
potential considered in \cite{burrill16}.  We find that the results from ZE-DOS agree reasonably
closely with ZE-$\bar{\mu}_e$ and the the results using $V^{AA}$ are larger than using $V^{PA}$.
In the center panel we compare the ZE-DOS results to those using the KG formulation.  For the
same potential the ZE and KG are reasonably close, but at lowest temperatures KG
with $V^{AA}$ gives a somewhat larger conductivity than ZE with the same potential.

In the bottom panel of figure \ref{fig_alsigff} we compare to other calculations and experiments.  First there
is an experiment due to Milchberg {\it et al} \cite{milchberg88} and a very recent
experiment due to Sperling {\it et al} \cite{sperling15}.  The experiments are in significant disagreement
with each other.  A recent simulation that post-processes an Orbital Free MD simulation with 
the Kohn-Sham KG method is also shown (Sjostrom {\it et al} \cite{sjostrom15}).  In the analysis
of reference \cite{sjostrom15} they demonstrate that their simulations agree with 
previous KS-DFT-MD results at lower temperatures as well as other experiments, casting
doubt on the veracity of the new measurements of Sperling {\it et al}.  
Finally we also show a calculation of Faussurier {\it et al} that uses the SCAALP
model \cite{faussurier15}.  This is a model that is similar in spirit to the PAMD
model used here, though differs in details.  Their conductivity calculation uses
the Ziman-Evans method and is qualitatively most similar to our calculations when we
use ZE-$\mu_e$ with $V^{AA}$.  Comparing the result of Faussurier {\it et al} to this
calculation (bottom panel) we find good agreement which provides gross check that our
calculations are reasonable.  

Now comparing our calculations using $V^{PA}$ with ZE-DOS or KG to the simulations of Sjostrom {\it et al}
we find that while they agree well with each other, they do not agree well the Sjostrom {\it et al} \cite{sjostrom15}, with our 
calculations predicting a significantly smaller conductivity.  Moreover our results are in agreement with the
recent experiment of Sperling {\it et al} but underestimate the older experiment of Milchberg {\it et al}.  
Our results using $V^{AA}$ with either ZE-DOS or KG also do not agree well with Sjostrom {\it et al} either,
though the magnitude is improved.  
In fact the best agreement with the Sjostrom simulation is found when we emulate the calculation
of Faussurier {\it et al} (i.e. ZE-$\mu_e$: $V^{AA}$).  The evidence put forward in Sjostrom {\it et al} \cite{sjostrom15}
suggests that the experiment of Sperling, and in turn our results using $V^{PA}$, are
too small.  This significant disagreement between these simulations and our $V^{PA}$ results
stands in stark contrast to the excellent agreement found for hydrogen and beryllium (figures
\ref{fig_h1} and \ref{fig_be1}).  

To explain why we find such a strong disagreement
in this case we recall the reasoning behind the choice of the electron scattering potential $V^{PA}$.
In the limit where the Born approximation is valid (i.e. when the scattered electron can be modeled
as a plane wave) the total scattering cross section for the plasma separates into individual
scattering cross sections for each ion in the plasma, which depend only on the scattering potential
at that site.  In the language of the PAMD model this scattering potential is $V^{PA}(r)$.  Thus, the obvious approximation
when ignoring multiple scattering effects, but wishing to allow strong scattering (i.e. beyond the Born
approximation) is to continue to assume that each site scatters the electrons independently, but treat
the scattering cross section using the t-matrix approach.  In the KG formula the t-matrix is equivalent to using the
calculated scattered wavefunctions.  Thus we can expect $V^{PA}$ to be a reasonable scattering potential
where ever multiple scattering effects can be ignored.  One test that would indicate the validity of this assumption is if the Born
calculation itself gives reasonable results.  In fact, as shown in \cite{burrill16, cpw15},
Born approximation based calculations using the ZE formula give reasonable results for the hydrogen
and Beryllium cases we have looked at here, whereas for aluminum the Born result is in gross error.  This
is thus the likely reason for the failure of the $V^{PA}$ based calculation for aluminum at the lower
temperatures.  We stress that it is not a necessary condition that the Born approximation be valid
for the present models to be accurate, rather that the multiple scattering effect be negligible.  Indeed
the t-matrix based calculations presented in \cite{burrill16} are in significantly better agreement
with the KS-DFT-MD result for hydrogen than corresponding Born approximation results.

As shown in figure \ref{fig_alsigff} , using $V^{AA}$ we get somewhat improved agreement with
the simulations of Sjostrom {\it et al}.  Unlike $V^{PA}$, $V^{AA}$ does not recover the
Born limit (leading to the poor results using this potential in figures \ref{fig_h1} and \ref{fig_be1}).
Using $V^{AA}$ may partially compensate for the breakdown of the weak multiple scattering
effect assumptions, but not in a controlled way.  Therefore it is difficult to know for which
cases (i.e. element, densities and temperatures) that using $V^{AA}$ will lead to improved
agreement.  In light of this we must be cautious in interpreting this somewhat improved agreement
as evidence that $V^{AA}$ leads to improved agreement in general at low temperature.

\section{Conclusions\label{sec_con}}
A new derivation of the Kubo-Greenwood conductivity for average atom models
is given.  The derivation improves upon the previous derivation of Johnson {\it et al} \cite{johnson}
by taking the ionic structure factor explicitly into account.  Calculations based
on the new expression, using the pseudoatom molecular dynamics model \cite{starrett15} to provide
the inputs, result in a significant improvement in agreement with KS-DFT-MD
simulations for the hydrogen and beryllium plasmas tested.  
We have also given and tested a version of the Ziman-Evans formula
for the resistivity that takes into account a non-free electron density of states.
This new expression removes the ambiguity as to the choice of chemical potential
and density of scattering electrons that was discussed in details in \cite{burrill16}.
The new Ziman-Evans formula also gives good agreement with the KS-DFT-MD results for hydrogen 
and beryllium.  We have also given the expression for the thermal
conductivity based on the Kubo-Greenwood formulation, and found much improved 
agreement with KS-DFT-MD results for dense hydrogen.

Calculations from the new Ziman-Evans and Kubo-Greenwood formulas for warm dense aluminum
found relatively poor agreement with {\it ab initio} simulations.  Possible reasons 
for this are discussed.

\section*{Acknowledgments}
This work was performed under the auspices of the United States Department of Energy under contract DE-AC52-06NA25396
and LDRD number 20150656ECR.

\appendix
\section{Aspects of the derivation of the KG formula}
When deriving equation (\ref{sigkg}) we are faced with an integral of the form
\begin{eqnarray}
I & \equiv &
\int d\hatk 
\int d\hatkp
S(|\bk -\bkp|)\nonumber \\
&& \times
Y_{l_1,m_1}(\hatk)
Y_{l_2,m_2}^*(\hatkp)
Y_{l_3,m_3}^*(\hatk)
Y_{l_4,m_4}(\hatkp)
\end{eqnarray}
which simplifies to
\begin{eqnarray}
I & = &
\int d\hatk 
S(k)
Y_{l_1,m_1}(\hatk)
Y_{l_3,m_1}^*(\hatk)\nonumber\\
&& \times
\delta_{l_2,l_4}
\delta_{m_2,m_4}
\delta_{m_1,m_3}
\end{eqnarray}
where $\delta$ is the Kronecker delta, and $k=\sqrt{k^2 + {k^\prime}^2- k\,k^\prime\cos\theta_k}$.
When $S(k) = 1$, the integrals in $I$ separate and the result simplifies to $ \delta_{l_2,l_4} \delta_{m_2,m_4}
\delta_{l_1,l_3} \delta_{m_1,m_3} $.  To eliminate the remaining sum of magnetic quantum number $m$
a useful relation is
\begin{equation}
\sum_{m=-l}^{l} m^2 |Y_{lm}(\hatk)|^2 = \frac{l(l+1)(2l+1)}{8 \pi } \sin^2\theta_k
\end{equation}

\bibliographystyle{unsrt}
\bibliography{phys_bib}

\begin{thebibliography}{10}

\bibitem{lambert11}
Flavien Lambert, Vanina Recoules, Alain Decoster, Jean Clerouin, and Michael
  Desjarlais.
\newblock On the transport coefficients of hydrogen in the inertial confinement
  fusion regime a).
\newblock {\em Physics of Plasmas (1994-present)}, 18(5):056306, 2011.

\bibitem{hu14}
S.~X. Hu, L.~A. Collins, T.~R. Boehly, J.~D. Kress, V.~N. Goncharov, and
  S.~Skupsky.
\newblock First-principles thermal conductivity of warm-dense deuterium plasmas
  for inertial confinement fusion applications.
\newblock {\em Phys. Rev. E}, 89:043105, Apr 2014.

\bibitem{kubo57}
Ryogo Kubo.
\newblock Statistical-mechanical theory of irreversible processes. i. general
  theory and simple applications to magnetic and conduction problems.
\newblock {\em Journal of the Physical Society of Japan}, 12(6):570--586, 1957.

\bibitem{greenwood58}
D.A. Greenwood.
\newblock The boltzmann equation in the theory of electrical conduction in
  metals.
\newblock {\em Proceedings of the Physical Society}, 71(4):585, 1958.

\bibitem{hanson11}
David~E. Hanson, Lee~A. Collins, Joel~D. Kress, and Michael~P. Desjarlais.
\newblock Calculations of the thermal conductivity of national ignition
  facility target materials at temperatures near 10 ev and densities near 10
  g/cc using finite-temperature quantum molecular dynamics.
\newblock {\em Physics of Plasmas}, 18(8), 2011.

\bibitem{desjarlais02}
M.~P. Desjarlais, J.~D. Kress, and L.~A. Collins.
\newblock Electrical conductivity for warm, dense aluminum plasmas and liquids.
\newblock {\em Phys. Rev. E}, 66:025401, Aug 2002.

\bibitem{sjostrom15}
Travis Sjostrom and J\'er\^ome Daligault.
\newblock Ionic and electronic transport properties in dense plasmas by
  orbital-free density functional theory.
\newblock {\em Phys. Rev. E}, 92:063304, 2015.

\bibitem{feynman}
R.~P. Feynman, N.~Metropolis, and E.~Teller.
\newblock Equations of state of elements based on the generalized fermi-thomas
  theory.
\newblock {\em Phys. Rev.}, 75:1561--1573, May 1949.

\bibitem{liberman}
David~A. Liberman.
\newblock Self-consistent field model for condensed matter.
\newblock {\em Phys. Rev. B}, 20:4981--4989, Dec 1979.

\bibitem{piron3}
R.~Piron and T.~Blenski.
\newblock Variational-average-atom-in-quantum-plasmas (vaaqp) code and virial
  theorem: Equation-of-state and shock-hugoniot calculations for warm dense al,
  fe, cu, and pb.
\newblock {\em Phys. Rev. E}, 83:026403, Feb 2011.

\bibitem{wilson}
B.~Wilson, V.~Sonnad, P.~Sterne, and W.~Isaacs.
\newblock Purgatorio—a new implementation of the inferno algorithm.
\newblock {\em Journal of Quantitative Spectroscopy and Radiative Transfer},
  99(1–3):658 -- 679, 2006.
\newblock Radiative Properties of Hot Dense Matter.

\bibitem{scaalp}
G\'erald Faussurier, Christophe Blancard, Philippe Coss\'e, and Patrick
  Renaudin.
\newblock Equation of state, transport coefficients, and stopping power of
  dense plasmas from the average-atom model self-consistent approach for
  astrophysical and laboratory plasmas.
\newblock {\em Physics of Plasmas}, 17(5), 2010.

\bibitem{rozsnyai}
Balazs~F. Rozsnyai.
\newblock Relativistic hartree-fock-slater calculations for arbitrary
  temperature and matter density.
\newblock {\em Phys. Rev. A}, 5:1137--1149, Mar 1972.

\bibitem{burrill16}
D.J. Burrill, D.V. Feinblum, M.R.J. Charest, and C.E. Starrett.
\newblock Comparison of electron transport calculations in warm dense matter
  using the ziman formula.
\newblock {\em High Energy Density Physics}, 19:1 -- 10, 2016.

\bibitem{faussurier15}
G\'erald Faussurier and Christophe Blancard.
\newblock Resistivity saturation in warm dense matter.
\newblock {\em Phys. Rev. E}, 91:013105, Jan 2015.

\bibitem{sterne07}
P.A. Sterne, S.B. Hansen, B.G. Wilson, and W.A. Isaacs.
\newblock Equation of state, occupation probabilities and conductivities in the
  average atom purgatorio code.
\newblock {\em High Energy Density Physics}, 3(1–2):278 -- 282, 2007.
\newblock Radiative Properties of Hot Dense Matter.

\bibitem{perrot87}
Fran\c{c}ois Perrot and M.~W.~C. Dharma-wardana.
\newblock Electrical resistivity of hot dense plasmas.
\newblock {\em Phys. Rev. A}, 36:238--246, Jul 1987.

\bibitem{pain10}
J.C. Pain and G.~Dejonghe.
\newblock Electrical resistivity in warm dense plasmas beyond the average-atom
  model.
\newblock {\em Contributions to Plasma Physics}, 50(1):39--45, 2010.

\bibitem{dharma06}
M.~W.~C. Dharma-wardana.
\newblock Static and dynamic conductivity of warm dense matter within a
  density-functional approach: Application to aluminum and gold.
\newblock {\em Phys. Rev. E}, 73:036401, Mar 2006.

\bibitem{rozsnyai08}
Balazs~F. Rozsnyai.
\newblock Electron scattering in hot/warm plasmas.
\newblock {\em High Energy Density Physics}, 4(1):64--72, 2008.

\bibitem{mazevet03}
S.~Mazevet, L.A. Collins, N.H. Magee, J.D. Kress, and J.J. Keady.
\newblock Quantum molecular dynamics calculations of radiative opacities.
\newblock {\em Astronomy \& Astrophysics}, 405(1):L5--L9, 2003.

\bibitem{johnson}
W.R. Johnson, C.~Guet, and G.F. Bertsch.
\newblock Optical properties of plasmas based on an average-atom model.
\newblock {\em Journal of Quantitative Spectroscopy and Radiative Transfer},
  99(1–3):327 -- 340, 2006.
\newblock Radiative Properties of Hot Dense Matter.

\bibitem{johnson09}
W.R. Johnson.
\newblock Low-frequency conductivity in the average-atom approximation.
\newblock {\em High Energy Density Physics}, 5(1–2):61 -- 67, 2009.

\bibitem{johnson2}
M.~Yu. Kuchiev and W.~R. Johnson.
\newblock Low-frequency plasma conductivity in the average-atom approximation.
\newblock {\em Phys. Rev. E}, 78:026401, Aug 2008.

\bibitem{starrett12a}
C.~E. Starrett, J~Cl{\'e}rouin, V~Recoules, J.~D. Kress, L.~A. Collins, and
  D.~E. Hanson.
\newblock Average atom transport properties for pure and mixed species in the
  hot and warm dense matter regimes.
\newblock {\em Physics of Plasmas (1994-present)}, 19(10):102709, 2012.

\bibitem{milchberg88}
H.~M. Milchberg, R.~R. Freeman, S.~C. Davey, and R.~M. More.
\newblock Resistivity of a simple metal from room temperature to ${10}^{6}$ k.
\newblock {\em Phys. Rev. Lett.}, 61:2364--2367, Nov 1988.

\bibitem{sperling15}
P.~Sperling, E.~J. Gamboa, H.~J. Lee, H.~K. Chung, E.~Galtier, Y.~Omarbakiyeva,
  H.~Reinholz, G.~R\"opke, U.~Zastrau, J.~Hastings, L.~B. Fletcher, and S.~H.
  Glenzer.
\newblock Free-electron x-ray laser measurements of collisional-damped plasmons
  in isochorically heated warm dense matter.
\newblock {\em Phys. Rev. Lett.}, 115:115001, 2015.

\bibitem{starrett15}
C.~E. Starrett, J.~Daligault, and D.~Saumon.
\newblock Pseudoatom molecular dynamics.
\newblock {\em Phys. Rev. E}, 91:013104, Jan 2015.

\bibitem{starrett13}
C.~E. Starrett and D.~Saumon.
\newblock Electronic and ionic structures of warm and hot dense matter.
\newblock {\em Phys. Rev. E}, 87:013104, Jan 2013.

\bibitem{evans73}
R.~Evans, B.~L. Gyorfey, N.~Szabo, and J.~M. Ziman.
\newblock On the resistivity of liquid transition metals.
\newblock In S.~Takeuchi, editor, {\em The properties of liquid metals}. Taylor
  and Francis, London, 1973.

\bibitem{ziman61}
J.M. Ziman.
\newblock A theory of the electrical properties of liquid metals. i: The
  monovalent metals.
\newblock {\em Philosophical Magazine}, 6(68):1013--1034, 1961.

\bibitem{potekhin96}
A.~Yu Potekhin and D.G. Yakovlev.
\newblock Electron conduction along quantizing magnetic fields in neutron star
  crusts. ii. practical formulae.
\newblock {\em Astronomy and Astrophysics}, 314:341--352, 1996.

\bibitem{starrett14}
C.E. Starrett and D.~Saumon.
\newblock A simple method for determining the ionic structure of warm dense
  matter.
\newblock {\em High Energy Density Physics}, 10:35 -- 42, 2014.

\bibitem{daligault16}
J\'er\^ome Daligault, Scott~D. Baalrud, Charles~E. Starrett, Didier Saumon, and
  Travis Sjostrom.
\newblock Ionic transport coefficients of dense plasmas without molecular
  dynamics.
\newblock {\em Phys. Rev. Lett.}, 116:075002, 2016.

\bibitem{dirac}
Paul~AM Dirac.
\newblock Note on exchange phenomena in the thomas atom.
\newblock {\em Proceedings of the Cambridge Philosophical Society}, 26:376,
  1930.

\bibitem{ksdt}
Valentin~V. Karasiev, Travis Sjostrom, James Dufty, and S.~B. Trickey.
\newblock Accurate homogeneous electron gas exchange-correlation free energy
  for local spin-density calculations.
\newblock {\em Phys. Rev. Lett.}, 112:076403, Feb 2014.

\bibitem{dulca00}
Lucian Dulca, John Banhart, and Gerd Czycholl.
\newblock Electrical conductivity of finite metallic systems: Disorder.
\newblock {\em Phys. Rev. B}, 61:16502--16513, 2000.

\bibitem{starrett15a}
C.E. Starrett.
\newblock A green's function quantum average atom model.
\newblock {\em High Energy Density Physics}, 16:18 -- 22, 2015.

\bibitem{cpw15}
D.~Burrill, D.~Feinblum, M.~Charest, and C.~Starrett.
\newblock Modeling warm dense matter using quantum-mechanical density function
  theory.
\newblock {\em Los Alamos National Laboratory Computational Physics Workshop
  2015 report}, 2015.

\end{thebibliography}

\end{document}